\documentstyle{article}

\def\be{\begin{equation}}
\def\ee{\end{equation}}
\def\bea{\begin{eqnarray}}
\def\eea{\end{eqnarray}}
\begin{document}

\pagestyle{empty}
\vskip-10pt
\hfill {\tt hep-th/0012070}

\begin{center}
\vskip 5truecm
{\LARGE \bf
Commutation relations for surface operators in six-dimensional $(2, 0)$ theory}\\ 
\vskip 2truecm
{\large \bf M{\aa}ns Henningson}\\
\vskip 1truecm
{\it Institute of Theoretical  Physics, Chalmers University of Technology\\
S-412 96 G\"{o}teborg, Sweden}\\
\vskip 5truemm
{\tt mans@fy.chalmers.se}
\end{center}
\vskip 2truecm
\noindent{\bf Abstract:}
The $A_{N - 1}$ $(2, 0)$ superconformal theory has an observable associated with every two-cycle in six dimensions. We make a natural guess for the commutation relations of these operators, which reduces to the commutation relations of Wilson and 't Hooft lines in four-dimensional $SU (N)$ $N = 4$ super Yang-Mills theory upon compactification on a two-torus. We then verify these commutation relations by considering the theory at a generic point of its moduli space and including in the surface operators only contributions from the light degrees of freedom, which amount to $N - 1$ $(2, 0)$ tensor multiplets.   
\newpage
\pagestyle{plain}

\section{Introduction}
Wilson and `t Hooft lines provide important non-local order parameters in $(3 + 1)$-dimensional Yang-Mills theories.  To be specific, we will consider the case with an $SU (N)$ gauge group, with all fields of the theory invariant under the ${\bf Z}_N$ center. (Our main interest is the $N = 4$ super Yang-Mills theory, where all fields transform in the adjoint representation of $SU (N)$, which is indeed invariant under the center.) The Wilson line $A (\gamma)$ associated with a closed curve $\gamma$ in three-dimensional space (at some fixed time $t$) is then given by $A (\gamma) = {\rm Tr} P \exp i \int_\gamma A_\mu d x^\mu$, where $A_\mu$ is the $SU (N)$ connection, $P$ denotes path ordering, and the trace is taken in the fundamental representation of $SU (N)$. The 't Hooft line $B (\gamma^\prime)$ associated with a closed curve $\gamma^\prime$ in space is given by a multi-valued $SU (N)$ gauge transformation that changes by the element $\exp 2 \pi i / N$ of the ${\bf Z}_N$ center of $SU (N)$ as one encircles a curve $\gamma$ in space that links $\gamma^\prime$ once. (Such a gauge transformation has to be singular along $\gamma^\prime$.) Despite these different looking definitions, the Wilson and 't Hooft lines are closely related and are in fact each others electric-magnetic duals. They enjoy the following equal time commutation relations:
\be
A (\gamma) B (\gamma^\prime) = B (\gamma^\prime) A (\gamma) \exp \frac{2 \pi i}{N} L (\gamma, \gamma^\prime), \label{line-relations}
\ee
where $L (\gamma, \gamma^\prime)$ denotes the linking number of the curves $\gamma$ and $\gamma^\prime$ in three-dimensional space \cite{tHooft}. Furthermore, all Wilson lines commute with each other, and all 't Hooft lines commute with each other.

It now seems that there might be a $(5 + 1)$-dimensional origin of $(3 + 1)$-dimensional Yang-Mills theory. In particular, $N = 4$ super Yang-Mills theory with gauge group $SU (N)$ arises when the $A_{N - 1}$ $(2, 0)$ superconformal theory is compactified on a two-torus \cite{9507121}\cite{9512059}. These $(5 + 1)$-dimensional theories are self-dual, and we expect that there is only a single type of non-local order parameter $W (\Sigma)$ associated with a two-cycle $\Sigma$ in five-dimensional space (at some fixed time $t$). However, our limited understanding of the theory does not permit us to give a precise definition of this operator. The relationship to the order parameters in $(3 + 1)$-dimensional Yang-Mills theory discussed above is that the Wilson line $A (\gamma)$ is given by a surface operator $W (\Sigma)$ with $\Sigma$ a product of $\gamma$ and the $a$-cycle of the two-torus. Similarly, the 't Hooft line $B (\gamma^\prime)$ is given by a surface operator $W (\Sigma^\prime)$ with $\Sigma^\prime$ a product of $\gamma^\prime$ and the $b$-cycle of the two-torus. Electric-magnetic duality, which interchanges the $a$- and $b$-cycles of the two-torus, then interchanges the Wilson and 't Hooft lines as required. Furthermore, there is an obvious guess for the equal time commutation relations of the surface operators that would reproduce (\ref{line-relations}), namely
\be
W (\Sigma) W (\Sigma^\prime) = W (\Sigma^\prime) W (\Sigma) \exp \frac{2 \pi i}{N} L (\Sigma, \Sigma^\prime) , \label{surface-relations}
\ee
where $L (\Sigma, \Sigma^\prime)$ now denotes the linking number of the two-cycles $\Sigma$ and $\Sigma^\prime$ in five-dimensional space. 

The object of the present letter is to verify (\ref{surface-relations}). Despite our limited understanding of the $(2, 0)$ theory in general, and the proper definition of the surface operator $W (\Sigma)$ in particular, this is indeed possible as we will now outline: The $A_{N - 1}$ $(2, 0)$ superconformal theory can be thought of as the world-volume theory on $N$ parallell five-branes in eleven-dimensional $M$-theory. At a generic point in the moduli space, where all the five-branes are separated, the light degrees of freedom are given by $N$ non-interacting $(2, 0)$ tensor multiplets, each comprising a chiral two-form $X$ (i.e. with self-dual field strength $d X = * d X$), an $Sp (4)$ symplectic Majorana-Weyl fermion $\psi$ and an $SO (5)$ vector $\phi^A$ of scalars that parametrize the transverse position of the five-brane in eleven dimensions. ($Sp (4) \simeq SO (5)$ is the $R$-symmetry of the $(2, 0)$ superconformal algebra.) However, a single tensor multiplet, describing the collective degrees of freedom of the $N$ five-brane system, is not really part of the internal world-volume theory and should be factored out. (The scalars of this multiplet describe the transverse position of the center of mass of the brane system.) This is in analogy with the corresponding situation for $N$ parallell three-branes in type IIB string theory, where the world-volume theory is a super Yang-Mills theory with gauge group $SU (N)$ rather than $U (N) \simeq SU (N) \times U(1)$. Returning to the $M$-theory configuration, we are thus left with $N - 1$ tensor multiplets, but these are not completely free since they inherit a subtle interaction from their origin on $N$ different five-branes. This point of view was advanced in \cite{0006231}, where it was shown how certain features of the full $(2, 0)$ tensionless string theory on space-times of non-trivial topology were reproduced already by this low-energy theory. In this letter, we will similarly show that including in the surface operator $W (\Sigma)$ only the contributions from these light degrees of freedom (which are known) is enough to reproduce (\ref{surface-relations}). This approach is fairly natural, since only massless degrees of freedom can be responsible for a long-range effect involving the linking number of two surfaces. 

As could be expected, the chiral two-forms in the tensor multiplets will play a crucial role for our argument. Already the quantum theory of a single such field is rather subtle, and it is by now generally accepted that it cannot be described by a covariant Lagrangian in the usual sense. Instead, the proper procedure is to factorize the theory of an ordinary two-form (which has a Lagrangian description) as the theory of a chiral two-form times the theory of an anti-chiral two-form \cite{9610234}\cite{9908107}. In this spirit, our approach to veryfing (\ref{surface-relations}) will be to first perform a canonical analysis of an ordinary two-form. The classical phase space factorizes into two parts pertaining to a chiral and anti-chiral two-form respectively, and the surface operator factorizes accordingly. It is therefore consistent to retain only the chiral part. We then take $N$ copies of the chiral part to build the phase space of $N$ non-interacting chiral two-forms and take the surface operator to be the product of the observables for the individual chiral two-forms. Finally, we factorize this observable into a part coupling to the chiral two-form representing the collective degrees of freedom and a part coupling to the remaining $N - 1$ chiral two-forms. The latter part is the one which defines the contribution to the surface operator $W (\Sigma)$ for the $A_{N - 1}$-type $(2, 0)$ theory. The commutation relations (\ref{surface-relations}) now follow from a canonical quantization procedure, where the classical Poisson bracket of functions on the classical phase space is replaced by the commutator of operators on the quantum Hilbert space.

\section{The phase space of a chiral two-form}
We work in $(5 + 1)$-dimensional Minkowski space with coordinates $x^\mu = (x^0, x^i) = (t, {\bf x})$, $\mu = 0, 1, \ldots, 5$, $i = 1, \ldots, 5$ and metric $\eta_{0 0} = -1$, $\eta_{i j} = \delta_{i j}$, $\eta_{0 i} = 0$, $i, j = 1, \ldots, 5$. 

Consider first an ordinary two-form $X_{\mu \nu} = - X_{\nu \mu}$ subject to the gauge invariance $X_{\mu \nu} \rightarrow X_{\mu \nu} + \Delta X_{\mu \nu}$, where the gauge parameter $\Delta X_{\mu \nu}$ is a closed two-form with integer periods. In Minkowski space, the Poincar\'e lemma states that $\Delta X_{\mu \nu}$ is exact and thus has zero periods, but in non-trivial topology we should also allow for `large' gauge transformations where $\Delta X_{\mu \nu}$ represents a non-trivial integral cohomology class. Given a two-cycle $\Sigma$, the `Wilson surface' observable $\exp 2 \pi i k \int_\Sigma X_{\mu \nu} d x^\mu \wedge d x^\nu$ is gauge invariant (even under `large' gauge transformations) for arbitrary integer $k$. If $\Sigma$ is the boundary of some open three-manifold $D$ (as is always the case in the topologically trivial Minkowski space), we may use Stokes' theorem to rewrite the `Wilson surface' as $\exp 2 \pi i k \int_D H_{\mu \nu \rho} d x^\mu \wedge d x^\nu \wedge d x^\rho$, where the gauge invariant field strength $H_{\mu \nu \rho}$ is given by $H_{\mu \nu \rho} = \partial_\mu X_{\nu \rho} + \partial_\nu X_{\rho \mu} + \partial_\rho X_{\mu \nu}$. Replacing $H_{\mu \nu \rho}$ by its dual $(* H)_{\mu \nu \rho} = \frac{1}{6} \epsilon_{\mu \nu \rho}{}^{\mu^\prime \nu^\prime \rho^\prime} H_{\mu^\prime \nu^\prime \rho^\prime}$ would instead give the dual `'t Hooft surface'.

The dynamics of the theory is governed by the Maxwell-type Lagrangian density 
\be
{\cal L} = - \frac{1}{6} H_{\mu \nu \rho} H^{\mu \nu \rho} .
\ee 
The canonical momenta $\Pi^{\mu \nu} = - \Pi^{\nu \mu}$ are defined as $\Pi^{\mu \nu} = \frac{\delta {\cal L}}{\delta \dot{X}_{\mu \nu}}$, where the dot indicates a derivative with respect to time $t$. We find that 
\bea
\Pi^{0 i} & = & 0 \cr
\Pi^{i j} & = & - H^{0 i j}.
\eea
The first of these equations is a first class constraint due to the gauge invariance, which we supplement with the temporal gauge condition $X_{0 i} = 0$. The second equation then reads 
\be
\Pi^{i j} = \dot{X}^{i j}.
\ee 
The Poisson bracket for the phase space variables $X_{i j}$ and $\Pi^{i j}$ (at equal time) is
\bea
\left\{ X_{i j} ({\bf x}), X_{i^\prime j^\prime} ({\bf x}^\prime) \right\} & = & 0 \cr 
\left\{ \Pi^{i j} ({\bf x}), \Pi^{i^\prime j^\prime} ({\bf x}^\prime) \right\} & = & 0 \cr
\left\{ X_{i j} ({\bf x}), \Pi^{i^\prime j^\prime} ({\bf x}^\prime) \right\} & = & \frac{1}{2} (\delta_i^{i^\prime} \delta_j^{j^\prime} - \delta_j^{i^\prime} \delta_i^{j^\prime}) \delta^{(5)} ({\bf x} - {\bf x}^\prime) .
\eea
The components of the field strength are $H_{0 i j} = \Pi_{i j}$ and $H_{i j k} = \partial_i X_{j k} + \partial_j X_{k i} + \partial_k X_{i j}$ with Poisson bracket
\bea
\left\{ H_{0 i j} ({\bf x}), H_{0 i^\prime j^\prime} ({\bf x}^\prime) \right\} & = & 0 \cr
\left\{ H_{i j k} ({\bf x}), H_{i^\prime j^\prime k^\prime} ({\bf x}^\prime) \right\} & = & 0 \cr
\left\{ H_{i j k} ({\bf x}), H_{0 i^\prime j^\prime} ({\bf x}^\prime) \right\} & = & \frac{1}{2} \left((\delta_{j i^\prime} \delta_{k j^\prime} - \delta_{j j^\prime} \delta_{k i^\prime}) \partial_i + (\delta_{k i^\prime} \delta_{i j^\prime} - \delta_{k j^\prime} \delta_{i i^\prime}) \partial_j + (\delta_{i i^\prime} \delta_{j j^\prime} - \delta_{i j^\prime} \delta_{j i^\prime}) \partial_k \right) \cr
& & \delta^{(5)} ({\bf x} - {\bf x}^\prime) .
\eea

We now decompose the field strength as $H_{\mu \nu \rho} = H^+_{\mu \nu \rho} + H^-_{\mu \nu \rho}$, where the self-dual and anti self-dual parts are given by
\bea
H^+_{\mu \nu \rho} & = & \frac{1}{2} \left(H_{\mu \nu \rho} + \frac{1}{6} \epsilon_{\mu \nu \rho}{}^{\mu^\prime \nu^\prime \rho^\prime} H_{\mu^\prime \nu^\prime \rho^\prime} \right) \cr
H^-_{\mu \nu \rho} & = & \frac{1}{2} \left(H_{\mu \nu \rho} - \frac{1}{6} \epsilon_{\mu \nu \rho}{}^{\mu^\prime \nu^\prime \rho^\prime} H_{\mu^\prime \nu^\prime \rho^\prime} \right) .
\eea
As independent components we may take $H^+_{i j k}$ and $H^-_{i j k}$ with Poisson bracket
\bea
\left\{ H^+_{i j k} ({\bf x}), H^-_{i^\prime j^\prime k^\prime} ({\bf x}^\prime) \right\} & = & 0 \cr
\left\{ H^+_{i j k} ({\bf x}), H^+_{i^\prime j^\prime k^\prime} ({\bf x}^\prime) \right\} & = & \frac{1}{4} \left(\epsilon_{i^\prime j^\prime k^\prime j k} \partial_i + \epsilon_{i^\prime j^\prime k^\prime k i} \partial_j + \epsilon_{i^\prime j^\prime k^\prime i j} \partial_k \right) \delta^{(5)} ({\bf x} - {\bf x}^\prime) \cr
\left\{ H^-_{i j k} ({\bf x}), H^-_{i^\prime j^\prime k^\prime} ({\bf x}^\prime) \right\} & = & - \frac{1}{4} \left(\epsilon_{i^\prime j^\prime k^\prime j k} \partial_i + \epsilon_{i^\prime j^\prime k^\prime k i} \partial_j + \epsilon_{i^\prime j^\prime k^\prime i j} \partial_k \right) \delta^{(5)} ({\bf x} - {\bf x}^\prime) . \label{PB}
\eea
The vanishing of the first of these expressions shows that the chiral and anti-chiral parts of the theory indeed decouple from each other, so it is consistent to consider only the former part with independent phase space variables $H^+_{i j k}$. 

\section{The commutation relations of surface operators}
In the chiral theory, the surface operators of Wilson and 't Hooft type are both given by 
\be
W_k (\Sigma) = \exp 2 \pi i k \int_D H^+_{\mu \nu \rho} d x^\mu \wedge d x^\nu \wedge d x^\rho ,
\ee
where again $\Sigma$ is the boundary of the three-manifold $D$. We wish to determine the commutation relations of this operator and an analogous operator $W_{k^\prime} (\Sigma^\prime)$ when $\Sigma^\prime$ is the boundary of another three-manifold $D^\prime$. We assume that both $D$ and $D^\prime$ are three-manifolds in space at some fixed time $t$ and begin by computing the commutator
\bea
\left[ \int_D H^+_{i j k} d x^i \wedge d x^j \wedge d x^k \right. & , & \left. \int_{D^\prime} H^+_{i^\prime j^\prime k^\prime} d x^{i^\prime} \wedge d x^{j^\prime} \wedge d x^{k^\prime} \right] \cr
& & = - \frac{i}{2 \pi} \int_\Sigma d x^i \wedge d x^j \int_{D^\prime} d x^{i^\prime} \wedge d x^{j^\prime} \wedge d x^{k^\prime} \epsilon_{i j i^\prime j^\prime k^\prime} \delta^{(5)} ({\bf x} - {\bf x}^\prime) \cr
& & = - \frac{i}{2 \pi} \Sigma \cdot D^\prime \cr
& & = - \frac{i}{2 \pi} L (\Sigma, \Sigma^\prime) .
\eea
where we have converted the Poisson bracket (\ref{PB}) to an operator commutator according to the usual rules of canonical quantization and used Stokes' theorem in the first line. In the second line, $\Sigma \cdot D^\prime$ denotes the intersection number of $\Sigma$ and $D^\prime$, and in the third line we have used the definition of the linking number $L (\Sigma, \Sigma^\prime)$ of $\Sigma$ and $\Sigma^\prime$. The value of the overall numerical coefficient should in principle follow from our previous formulas if we had been more careful with normalizations of various quantities. To verify that the value we have given here (and which is necessary for the sequel of our argument) is indeed the correct one would be a reassuring and instructive, although not completely trivial, excercise. It now follows from the Baker-Hausdorff formula that
\be
W_k (\Sigma) W_{k^\prime} (\Sigma^\prime) =  W_{k^\prime} (\Sigma^\prime) W_k (\Sigma) \exp 2 \pi i k k^\prime  L (\Sigma, \Sigma^\prime) ,
\ee
i.e. $W_k (\Sigma)$ and $W_{k^\prime} (\Sigma^\prime)$ commute for integer $k$ and $k^\prime$.

We now consider the case of $N$ non-interacting chiral two-forms $X^1_{\mu \nu}, \ldots, X^N_{\mu \nu}$ as is appropriate for $N$ widely separated five-branes in $M$-theory. It is convenient to assemble their field strengths into an $N$-vector ${\bf H}^+_{\mu \nu \rho} = (H^{1 +}_{\mu \nu \rho}, \ldots, H^{N +}_{\mu \nu \rho})$. The appropriate surface operators are of the form
\be
W_{\bf k} (\Sigma) = \exp 2 \pi i {\bf k} \cdot \int_D {\bf H}^+_{\mu \nu \rho} d x^\mu \wedge d x^\nu \wedge d x^\rho ,
\ee
where ${\bf k} = (k_1 , \ldots , k_N)$ and the raised dot denotes the standard scalar product of $N$-vectors. Their commutation relations are
\be
W_{\bf k} (\Sigma) W_{\bf k^\prime} (\Sigma^\prime) = W_{\bf k^\prime} (\Sigma^\prime) W_{\bf k} (\Sigma) \exp 2 \pi i {\bf k} \cdot {\bf k^\prime} L (\Sigma, \Sigma^\prime) ,
\ee
so that $W_{\bf k} (\Sigma)$ and $W_{\bf k^\prime} (\Sigma^\prime)$ commute for ${\bf k}, {\bf k^\prime} \in {\bf Z}^N$.

Finally, we wish to separate the collective and internal world-volume degrees of freedom of the $N$ five-brane system. To this end, we decompose ${\bf k}$ as ${\bf k} = \kappa {\bf 1} + {\bf w}$, where $\kappa \in \frac{1}{N} {\bf Z}$, ${\bf 1} = (1, \ldots, 1)$ and ${\bf w}$ is an element of the weight lattice $\Gamma^w$ of $SU (N)$ defined as
\be
\Gamma^w = \left\{ (w_1, \ldots, w_N) \in (\frac{1}{N} {\bf Z})^N; w_I - w_J \in {\bf Z}, w_1 + \ldots + w_N = 0 \right\} .
\ee
The surface operator factorizes accordingly as $W_{\bf k} (\Sigma) = W_{\kappa \bf 1} (\Sigma) W_{\bf w} (\Sigma)$, where the first factor and second factor involves the collective and internal world-volume degrees of freedom respectively. The commutation relations of the latter are
\be
W_{\bf w} (\Sigma) W_{\bf w^\prime} (\Sigma^\prime) = W_{\bf w^\prime} (\Sigma^\prime) W_{\bf w} (\Sigma) \exp 2 \pi i {\bf w} \cdot {\bf w^\prime} L (\Sigma, \Sigma^\prime) , \label{w-rela}
\ee
but ${\bf w} \cdot {\bf w^\prime}$ is in general an element of $\frac{1}{N} {\bf Z}$ and not necessarily of ${\bf Z}$, so these operators need not commute.

In fact, if we introduce also the $SU (N)$ root lattice $\Gamma^r$ defined by 
\be
\Gamma^r = \left\{ (r_1, \ldots, r_N) \in {\bf Z}^N; r_1 + \ldots + r_N = 0 \right\},
\ee
which is dual to and a sublattice of the weight lattice $\Gamma^w$, we may decompose $\Gamma^w$ in cosets with respect to $\Gamma^r$. The phase factor $\exp 2 \pi i {\bf w} \cdot {\bf w}^\prime  L(\Sigma, \Sigma^\prime) $ only depends on ${\bf w}$ and ${\bf w}^\prime$ through their coset classes $[{\bf w}]$ and $[{\bf w}^\prime]$ in $\Gamma^w / \Gamma^r \simeq {\bf Z}_N$ and may therefore be written as $\exp 2 \pi i [{\bf w}] \cdot [{\bf w}^\prime] L(\Sigma, \Sigma^\prime)$. If we now take $[{\bf w}]$ and $[{\bf w}^\prime]$ to be the fundamental conjugacy classes, this phase factor equals $\exp \frac{2 \pi i}{N} L(\Sigma, \Sigma^\prime)$, and we reproduce (\ref{surface-relations}).

As a final remark, we note that a `large' gauge transformation (in a space of non-trivial topology) is given by a vector $\Delta {\bf X}_{\mu \nu} = (\Delta X_{\mu \nu}^1, \ldots, \Delta X_{\mu \nu}^N)$ of closed two-forms whose periods take their values in ${\bf Z}^N$. Those $\Delta {\bf X}_{\mu \nu}$ for which the periods take their values in the root lattice $\Gamma^r \subset {\bf Z}^N$ are the ones that do not affect the collective but only the internal world-volume degrees of freedom.

\vspace*{5mm}
This research was supported by the Swedish Natural Science Research Council (NFR).


\begin{thebibliography}{99}
\bibitem{tHooft}G.~'t~Hooft, `On the phase transition towards permanent quark confinement', {\it Nucl. Phys.} {\bf B138} (1978) 1.
\bibitem{9507121}E.~Witten, `Some comments on string dynamics', {\it Los Angeles 1995, Future perspectives in string theory} 501, {\tt hep-th/9507121}.
\bibitem{9512059}A.~Strominger, `Open $p$-branes', {\it Phys. Lett.} {\bf B383} (1996) 44, {\tt hep-th/9512059}.
\bibitem{0006231}M.~Henningson, `A class of six-dimensional conformal field theories', {\it Phys. Rev. Lett.} {\bf 85} (2000) 5280, {\tt hep-th/0006231}.
\bibitem{9610234}E.~Witten, `Five-brane effective action in $M$-theory', {\it J. Geom. Phys.} {\bf 22} (1997) 103, {\tt hep-th/9610234}.
\bibitem{9908107}M.~Henningson, B.E.W.~Nilsson, and P.~Salomonson, `Holomorphic factorization of correlation functions in $(4 k + 2)$-dimensional $(2 k)$-form gauge theory', {\it JHEP} {\bf 9909} (1999) 008, {\tt hep-th/9908107}.
\end{thebibliography}
\end{document}